\documentstyle[11pt,newpasp,twoside,epsf]{article}
\def\edcomment#1{\iffalse\marginpar{\raggedright\sl#1\/}\else\relax\fi}
\reversemarginpar

\begin{document}

\title{Detection of an inner torus in the Proto planetary Nebulae OH231.8+4.2}
\author{J.-F. Desmurs,  C. S\'anchez Contreras, V. Bujarrabal, F. Colomer \& J. Alcolea}
\affil{Observatorio Astron\'omico Nacional (IGN)}

\begin{abstract}

We performed the first VLBI observations of the SiO $v$=1 and $v$=2
$J$=1--0 masers in a Proto Planetary Nebula OH 231.8+4.2 (also known as
the Rotten Egg nebula), at milliarcsecond resolution. Only the $v$=2
maser transition was detected. We detect several maser spots lying
along a line which is almost perpendicular to the axis of symmetry of
the Nebula.  We find that all the emission is concentrated in an area
of $\sim$ 10~mas, indicating that the SiO masers are originated very
close to the surface of the star. One of the two emission areas
presents an elongated structure with a clear velocity gradient. The
detected emission is consistent with a torus, or disk, in rotation with
a velocity of $\sim$\,6 km/s and with an infall velocity of $\sim$ 10
km/s.

\end{abstract}

\section{Introduction}

\begin{figure}[t]
\plotfiddle{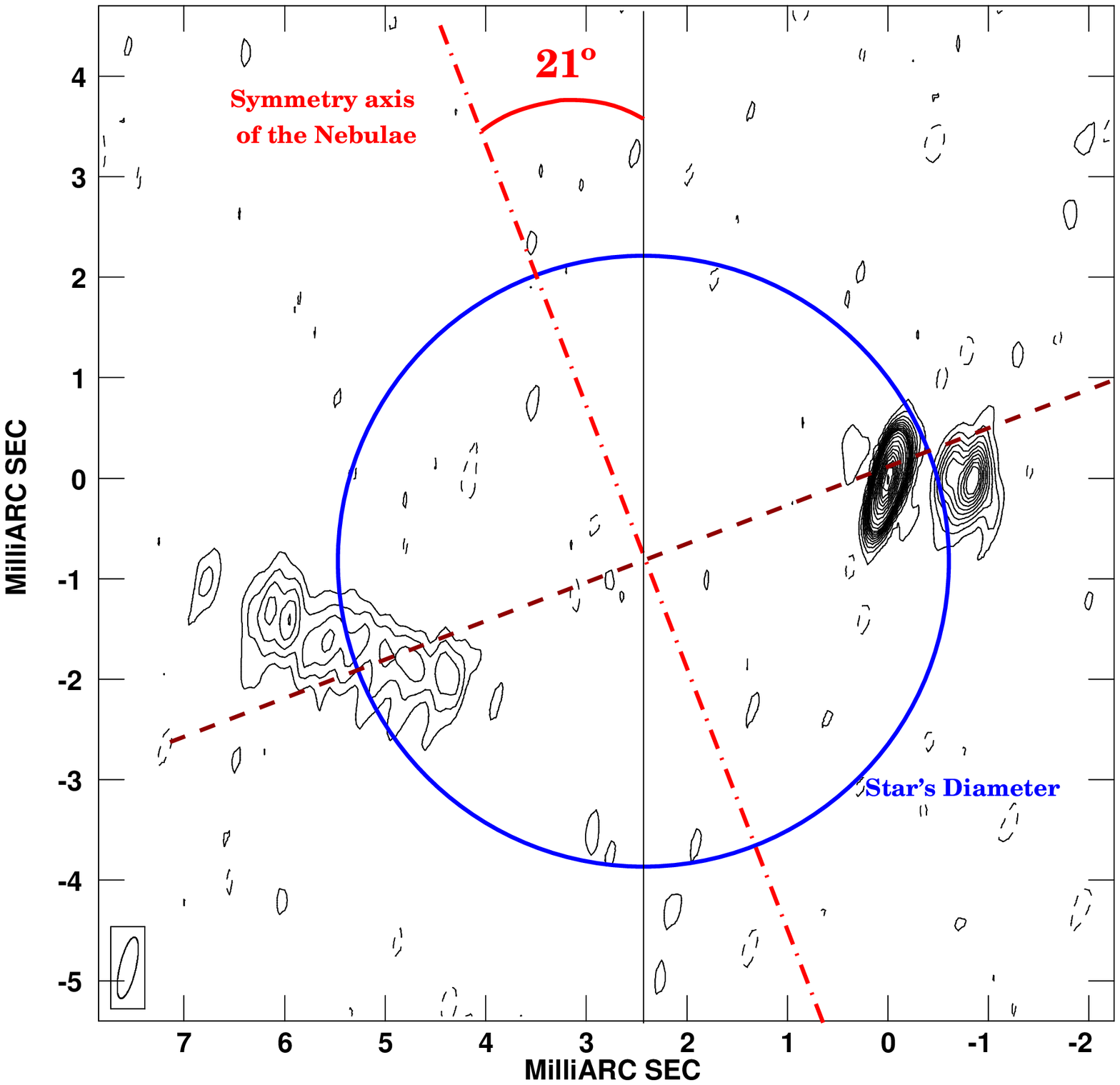}{5cm}{0}{50}{50}{-150}{-180}
\plotfiddle{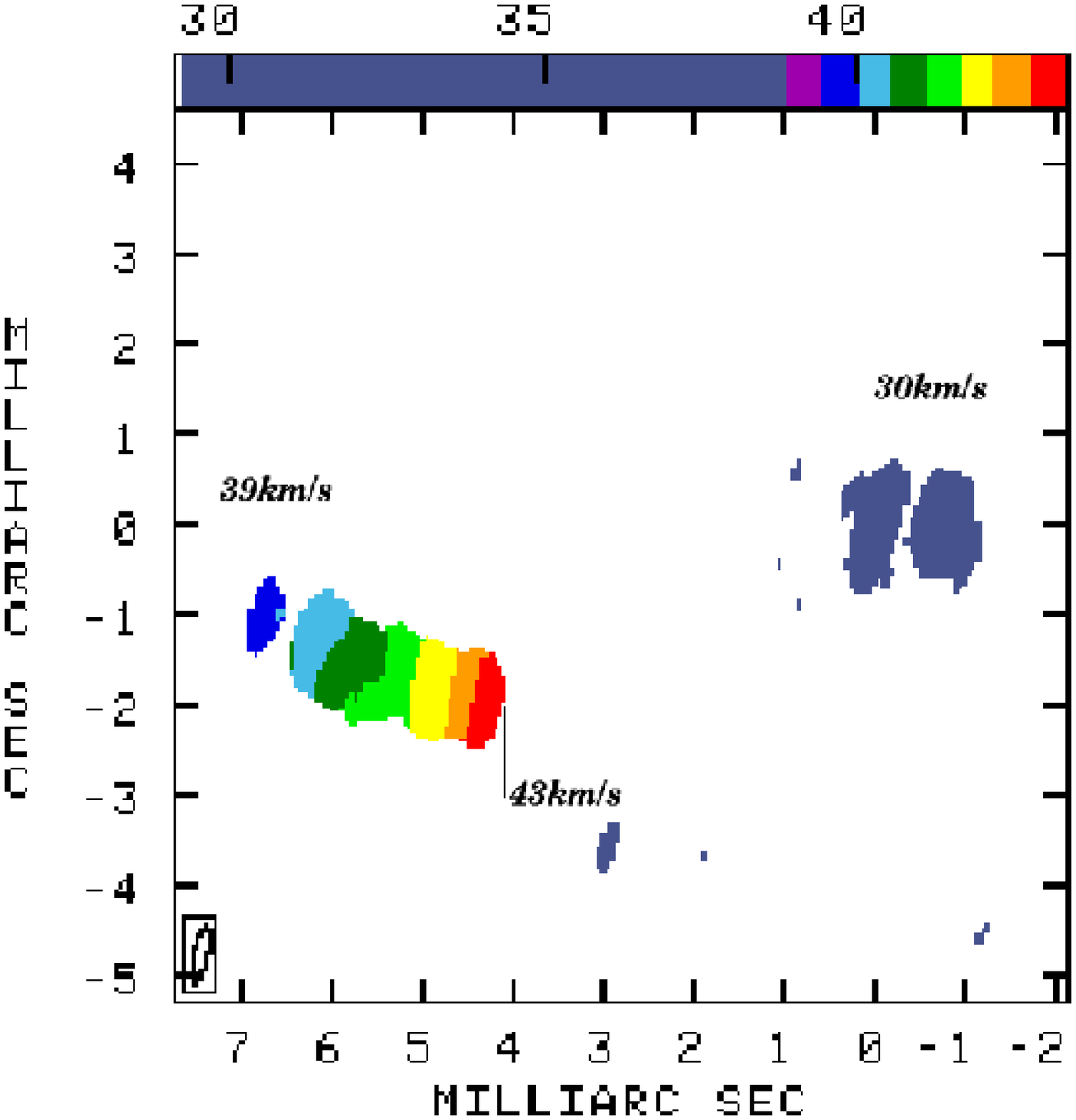}{13cm}{0}{53}{53}{-170}{-70}
\caption
{ \sl Top, the integrated flux map of OH 231.8+4.2 for $v$=2
$J$=1--0 SiO maser transition. Bottom, the integrated velocity map
showing the linear velocity gradient in the Eastern component. The
systemic velocity of the source is 33km/s.} 
\end{figure}

The evolution of the envelopes around AGB stars toward Planetary
Nebulae (PNe), through the stage of Proto Planetary Nebulae (PPNe) is
still yet poorly known. Contrarily to AGB envelopes that show a
spherical symmetry, Planetary and Proto Planetary Nebulae present a
conspicuous axial symmetry and very collimated jets. To explain their
evolution several models (see Mellema 1995, Soker 1998a,b, and
references therein) have postulated the presence of a dense ring or
disk close to the central post-AGB object. Such a disk would have a
double effect, on the one hand produces a mechanical collimation of the
ejections, and on the other hand allows a mechanism for the accretion
of material by the star.
 
The most popular of these models assume the presence of a dense disk or
ring, very close to the star, that would collimate the stellar wind in
the direction perpendicular to the disk plane.  Other models also
assume a central disk, but related to the accretion of circumstellar
material by the star, either by providing an orbiting, long-lived
reservoir of matter or/and acting as an accretion disk. In this case,
the fast bipolar jets appear by interaction of the infalling material
with the stellar magnetic field (a process similar to that explaining
the outflows in forming stars).  These theories both explain the very
collimated post-AGB jets and the axial symmetry of PPNe. Moreover,
accretion of circumstellar material explains the peculiar abundances
found in the atmospheres of certain post-AGB stars, that would be
affected by the composition of the reaccreted material (e.g$.$ Van
Winckel et al$.$ 1995), and the very high energy and momentum of the
bipolar outflows in PPNe, that cannot be explained by radiation
pressure and seem to require a release of gravitational energy
(Bujarrabal et al$.$ 1998).

Existing observations reveal the presence of central disks in several
PPNe, but their spatial resolution only allows the study of extended
regions, compared to the very inner regions of the disks that are
relevant for the above processes. Fortunately, SiO masers are detected
in one of the PPNe, OH231.8+4.2 (see Nyman et al$.$ 1998), allowing
VLBI observations of the close surroundings of the star.

OH 231.8+4.2 is a very well studied PPNe (see S\'anchez Contreras et
al$.$ 1997, 2000), having less than 1200 years old. It is situated at a
distance of $\sim$ 1500 pc and its luminosity is $\sim 10^4
L_{\odot}$. It is composed, as far as we know, by at least one evolved
star surrounded by a remarkable bipolar nebula. The symetry axis of the
nebula, at P.A. = 21$^o$, is inclined 36 $^o$ with respect to the plane
of the sky (the North lobe is pointing toward us).  The central core is
a cold star (M9\,III, $\sim 2000^o$ K) with a stellar radius of about
4.5 A.U deduced from its brightness temperature.  
Between the different PPNe known, it is one of the only one showing SiO
maser emission allowing line VLBI observation at sub milliarcsecond
resolution. 

\section{Observations and discussion}
We performed observations of the $v$=1 and $v$=2 $J$=1--0 lines of SiO
at 43.122080 and 42.820587 GHz with the NRAO Very Long Baseline Array
in May 2000. The system was setup to record 4MHz of left and right
circular polarizations in both lines. The correlation was produced at
the VLBA correlator in Socorro (NM, USA) providing 128 spectral
channels of the parallel and cross-hand polarization bands, achieving a
spectral resolution of 0.22 km/s, and giving access to the four Stokes
parameters and hence the linear polarization.

The data have been analyzed using the AIPS package in the standard
way. Only maps of the $v$=2 line have been produced as the $v$=1 line
was not detected (The $v$=2 line is known to be slightly stronger see
Jewell et al$.$ 1991, Nyman et al$.$ 1998).

In Figure 1, Top part, we show the integrated flux map of the $v$=2
$J$=1--0 line of SiO. The resolution beam of our observation is 0.6 by
0.16 mas (PA --15$^o$) and the noise level in the map is less
than 50 mJy. The integrated peak flux is 35 Jy/beam.  We map a square
region of about 50 mas but we found that all the emission was
concentrated in a small area of only few mas (up to 10 mas).  The blue
circle drawn on the map represent the diameter, at the map scale, of
the central Mira variable star ($\sim$ 9 A.U.), and the red lines
represent the symmetry axis of the bipolar Nebula in the plane of the
sky.

All the detected emission arises from two regions. That in the western
part of the map shows two strong components, while on the eastern side
of the map a linear structure roughly orientate South-West to
North-East.
\clearpage 
The whole emission is lying on a direction perpendicular to the
symmetry axis of the Nebula. Another important geometrical remark is
that the separation between the two emission structures is about 10
A.U.  Considering the star's diameter, this means that the detected
emission is originated very close to the star, nearly at its surface.

The velocity analysis of our data shows the presence of a clear
velocity gradient (Fig. 1, bottom): The projected velocity ranges from
43~km/s in the Southern spot down to 39~km/s in the extrem East part.
The other emission region in the West does not show any velocity
structure. Both emission spots appear at a fixed position and present a
broad line width of nearly 2 km/s.

The analysis and interpretation of the observed velocities are not
trivial. In the case of disk in a pure Keplerian rotation we would
expect to observe the highest velocities on one edge of the detected
torus/disk and to observe velocities close to systemic velocity on the
line of sight of the center of the torus/disk. In our observations, we
observed exactly the opposite behavior, with the redest velocities
close to the line of sight of the center.

To explain this, we have postulated that the systemic velocity is
combined with an infalling velocity. From our observations, we deduced
that the needed infall velocity must be of the order of 10 km/s.  The
fact that the Western (eastern) side of the detected torus/disk is red
(blue) shifted with respect to the systemic velocity of the source
($\sim$\,33~km/s)indicate that the disk is rotating clockwise.  The
rotation velocity can be extracted from the velocities observed on the
edges of the structures. As the disk (or torus) might be bigger than
the region emiting SiO masers this velocity can be only a lower
limit. From the red shifted emission, we concluded that the rotation
velocity is at least 6 km/s.

Applying these previous observational values to a very simplistic model
considering a disk in rotation and infall, and having an inclination
of 36$^o$ with the plane of the sky, we have been able to fit quite
well our observations and to reproduce well the velocity gradient of
the elongated structure.

\end{document}